\documentclass{optica-article}

\journal{opticajournal} 

\articletype{Research Article}

\begin{document}

\title{A Two-Stage Physically Grounded Optical Design Framework with Application to a Multi-Channel Focal Reducer}

\author{M. R. Najera,\authormark{1,*} J. Herrera,\authormark{1} Esteban Luna,\authormark{1} and Erika Sohn\authormark{1}}

\address{\authormark{1}Instituto de Astronomía UNAM (IA-E), Carr. Tijuana-Ensenada km107, CICESE, 22860 Ensenada, BC}

\email{\authormark{*}mnajera@astro.unam.mx} 


\begin{abstract*} 
We present a two-stage optimization framework that formalizes classical optical design strategies within a reproducible computational workflow. Starting point designs are constructed from paraxial and third-order models, providing optically consistent initial configurations. These configurations are refined through a physically grounded residual merit function evaluated by exact ray tracing, followed by RMS-based image refinement. Applied to a multi-channel astronomical focal reducer, the methodology provides a structured transition from aberration-based modeling to numerical optimization while preserving physically interpretable intermediate stages. By embedding classical optical design principles into an open computational pipeline, the approach reduces reliance on heuristic initial conditions and improves the reproducibility and interpretability of the design process.
\end{abstract*}


\section{Introduction}

  The design of optical systems is inherently a high-dimensional and nonlinear problem in which the final performance depends critically on the selection of initial configurations and the formulation of appropriate merit functions. Conventional optical design workflows are commonly driven by image-quality metrics, such as root mean square (RMS) spot size, wavefront error, modulation transfer function (MTF), Strehl ratio, and encircled energy (EE), evaluated through numerical ray tracing. These approaches have proven highly effective and constitute standard practice in modern optical design. However, because the optimization is typically formulated in terms of final image-plane performance, the intermediate physical relationships between design variables, aberration control, and system behavior are not always represented explicitly within the optimization process. To address this aspect, the present work introduces a Physically Grounded Merit Function (PGMF) as an intermediate residual-based optimization layer that complements conventional image-quality optimization by explicitly incorporating physically motivated constraints derived from classical optical design principles. The proposed framework is implemented within the open optical simulation library \texttt{KrakenOS}\cite{herrera2022krakenos}, and combines paraxial and third-order optical modeling with exact ray tracing, the PGMF optimization stage, and subsequent numerical refinement within a reproducible computational environment. Rather than introducing new aberration theory, the contribution of this work lies in the formalization of classical optical design strategies into an explicit computational workflow that preserves physically interpretable intermediate stages throughout the optimization process. The methodology is applied to the design of a refractive focal reduction system for a multi-channel astronomical instrument,\cite{najera2026opticaldesignopticamargthreechannel}, where stringent requirements on image quality, spectral coverage, and system compactness must be satisfied simultaneously.

  The remainder of this paper is organized as follows. Section~\ref{Sec:LimitationMF} motivates the proposed framework by discussing limitations of conventional merit-function formulations. Section~\ref{Sec:PGMF} introduces the Physically Grounded Merit Function, while Section~\ref{Sec:Methodology} describes the complete design workflow. Section~\ref{Sec:Results} presents the resulting optical designs and their performance evaluation. Finally, Sections~\ref{Sec:Discussion} and \ref{Sec:Conclusions} discuss the implications, limitations, and main conclusions of the proposed approach.

\section{Limitations of Conventional Merit Functions}\label{Sec:LimitationMF}

    Conventional optical design workflows, previously described as being driven by image-based metrics, are typically implemented through the minimization of weighted merit functions combining multiple performance indicators, each representing a specific design objective. While this approach provides considerable flexibility, it also introduces challenges affecting the interpretability, reproducibility, and physical transparency of the optimization process. A primary challenge lies in the selection of weighting factors, which are generally chosen heuristically and often require iterative tuning, making the final solution sensitive to design choices and reducing reproducibility. Another limitation arises from the fact that commonly used image-quality metrics are primarily evaluated at the image plane. Although these quantities provide physically meaningful measures of system performance, the physical constraints governing image formation are often incorporated indirectly through the optimization objectives. Consequently, the connection between design variables and optical principles such as Fermat’s principle or the Coddington equations may become less explicit during the optimization process. As a result, numerical performance can improve without necessarily providing direct insight into how individual parameters contribute to aberration control.

    These limitations became increasingly significant with the advent of computational ray tracing, as optical design evolved toward automated workflows in which the traditional multistage process based on paraxial analysis, third-order aberration control, and progressive refinement is often condensed into a single optimization routine. Recent advances have combined gradient-based optimization with global exploration strategies to identify suitable starting configurations and system architectures \cite{Cote:21,Zoric:24,Zoric:25}. However, these approaches still depend on expert intervention for tasks such as topology changes and the selection of initial configurations, since structural modifications require iterative refinement and manual guidance \cite{2025arXiv250923572T}. This limitation is closely related to the highly nonconvex nature of optical design problems, where multiple local minima may exist and the final solution can depend strongly on the initial configuration. Previous studies have shown that the optical design space exhibits a structured organization of solutions \cite{van2009finding}, with intrinsic relationships between configurations of different complexity that can be exploited for systematic exploration \cite{hou2015reducible}. Nevertheless, conventional merit functions are generally formulated in terms of image-quality objectives and numerical performance metrics, so the physical constraints associated with aberration control are often incorporated indirectly through the optimization targets rather than being represented explicitly as optimization residuals.

    Historically, lens design relied on physically motivated strategies that produced canonical configurations such as the Gauss doublet and the Cooke triplet, later formalized through aberration theory by Conrady and Kingslake \cite{conrady1957applied, conrady1960applied, kingslake1978lens}. Subsequent developments combined paraxial and third-order modeling with iterative optimization techniques \cite{sasian2019introduction, bentley2012field}, while comprehensive references consolidated optical design principles \cite{shannon1997art, malacara2003handbook, kidger2001fundamental}. Although automated optimization significantly increased design efficiency, many of the intermediate stages traditionally used in classical optical design became less visible within modern numerical workflows. The proposed approach does not seek to replace modern optical design principles, but rather to recover part of this physical interpretability within a reproducible computational environment by integrating paraxial analysis, aberration balancing, and intermediate optical models with exact ray tracing and modern numerical methods.

\section{Physically Grounded Merit Function}\label{Sec:PGMF}

    Within this framework, the physically grounded merit function (PGMF) connects classical optical models with exact ray-tracing-based optimization. The PGMF formulates quantities traditionally used in aberration analysis as residuals computable through exact ray tracing, establishing a direct connection between design variables and aberration control. Section~\ref{subSec:TheoryFun} introduces the physical conditions governing the primary aberrations of the system, while Section~\ref{subSec:MFformu} presents the corresponding merit function formulation.

    \subsection{Theoretical Foundations}\label{subSec:TheoryFun}
    
        The formulation of the PGMF is based on a reduced set of conditions that capture the dominant aberration behavior of the system through four primary components: optical path consistency derived from Fermat’s principle, chromatic path compensation, transverse coma control, and astigmatism correction through sagittal and tangential focus separation. Spherical aberration is addressed through Fermat’s principle by enforcing equality between the optical path lengths of marginal and principal rays,
        \begin{equation}
        \delta l = \overline{\mathrm{TOP}}_{m} - \overline{\mathrm{TOP}}_{p} \approx 0,
        \end{equation}
        where $\overline{\mathrm{TOP}}$ denotes the total optical path accumulated along the ray trajectory. Chromatic aberration is incorporated through variations in optical path length across multiple wavelengths. Using a reference wavelength $W_1$, the relative differences with respect to additional wavelengths are defined as
        \begin{equation}
        r_a = \delta l_{w_1} - \delta l_{w_2}, \qquad
        r_b = \delta l_{w_1} - \delta l_{w_3},
        \end{equation}
        which are combined into the residual
        \begin{equation}
        \delta r = \sqrt{r_a^2 + r_b^2} \approx 0.
        \end{equation}
                
        Coma and astigmatism are characterized through geometric quantities obtained from ray tracing. Coma is quantified using a transverse formulation that avoids explicit computation of the exit pupil,
        \begin{equation}
        \delta C_T = H_{12} - H_p \approx 0,
        \end{equation}
        providing a geometric interpretation of coma as a variation in lateral magnification. Astigmatism is described through the separation between sagittal and tangential focal positions obtained from ray intersections in orthogonal planes,
        \begin{equation}
        \delta f = |f_t - f_s| \approx 0,
        \end{equation}
        where $f_t$ and $f_s$ denote the tangential and sagittal focal distances. This formulation is consistent with the physical interpretation provided by the Coddington equations while remaining compatible with exact ray tracing.
    
    \subsection{Merit Function Formulation}\label{subSec:MFformu}
    
        Within the proposed framework, the PGMF is defined as a vector of residuals that enforce constraints derived from the optical behavior of the system. Each component captures a specific aspect of aberration control or system consistency. For a system defined by the design variables $\mathbf{x}$, the PGMF is given by
        \begin{equation}
        \mathbf{f}(\mathbf{x}) =
        \bigl(
        \delta l,\,
        w_r\,\delta r,\,
        \delta C_T,\,
        w_f\,\delta f,\,
        \Delta \mathrm{EFFL}
        \bigr),\label{eq:PGMF}
        \end{equation}
        where $\delta l$ enforces optical path consistency, $\delta r$ quantifies chromatic compensation, $\delta C_T$ represents the transverse coma residual, and $\delta f$ corresponds to the astigmatic focal separation. The term $\Delta \mathrm{EFFL}$ imposes a constraint on the effective focal length. Although the PGMF is formulated as a residual-based merit function, its distinction from conventional image-based merit functions lies in the physical interpretation of the optimized quantities, which are directly associated with aberration control and system consistency. The proposed framework therefore does not introduce a fundamentally different optimization mechanism. Rather, it reformulates the optimization problem in terms of physically interpretable residuals that explicitly represent aberration control conditions. In this way, the primary contribution of the PGMF lies in improving the interpretability of the intermediate optimization stage and providing physically meaningful guidance before the final image-based refinement.
                
       The reduced number of residual terms enables a controlled weighting strategy in which each coefficient has a well-defined role. The weighting factors $\omega_r$ and $\omega_f$ regulate the relative contribution of chromatic compensation and astigmatic correction, respectively, allowing balanced enforcement of the dominant aberrations without arbitrary scaling. The formulation can be extended to multiple field positions by incorporating both on-axis and off-axis contributions, ensuring uniform performance across the field of view. For more complex systems, additional field points can be included, yielding an expanded residual vector that preserves consistency between the number of constraints and the dimensionality of the design space. The optimization problem is then formulated as the minimization of a scalar objective function constructed from the residual vector,
        \begin{equation}
        \min_{\mathbf{x}} \sum_{i} \rho\left(f_i(\mathbf{x})^2\right),\label{eq:Equation2OPT}
        \end{equation}
        where $\rho(\cdot)$ denotes a suitable loss function. When $\rho(s)=s$, the formulation reduces to a standard least-squares (LS) problem. This structure allows the optimization to be defined independently of the numerical method while preserving a direct connection between the residuals and the optical behavior of the system.

\section{Methodology}\label{Sec:Methodology}

\begin{figure}[ht]
        \centering
        \includegraphics[width=0.36\columnwidth]{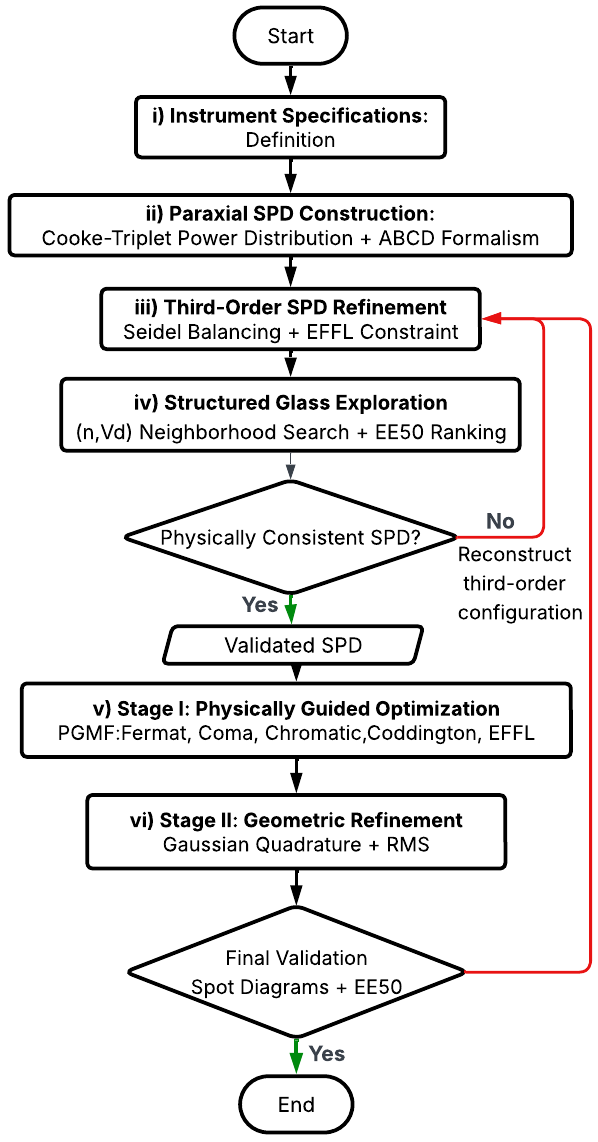}
        \caption{ 
        Conceptual workflow of the proposed two-stage optical design methodology.
        }
        \label{fig:OverallFC}
    \end{figure}

    The implementation of the proposed framework combines paraxial and third-order optical modeling with numerical ray tracing within a structured optimization workflow initialized from optically consistent starting point designs (SPD). The methodology is applied to the design of a refractive focal reduction system for a multi-channel astronomical instrument consisting of three independent channels defined by dichroic beam splitting, each incorporating a dedicated multi-element lens system that reduces the effective focal length of an f/8.5 telescope while maintaining image quality across an $8.4' \times 8.4'$ field of view. The design is constrained by instrumental specifications, including an $11\,\mu$m pixel size, a spectral range of $0.35$--$1.0\,\mu$m, representative seeing conditions ($\sim1.5''$), and system geometry \cite{najera2026opticaldesignopticamargthreechannel}. The resulting configuration yields a plate scale of $22.6\,''\,\mathrm{mm}^{-1}$ at the detector plane, corresponding to a sampling of approximately $0.25''$ per pixel, such that typical seeing is sampled over $\sim6$ pixels, indicating an oversampled regime. Figure~\ref{fig:OverallFC} summarizes the conceptual sequence of the methodology, from the definition of instrumental specifications to the construction of physically consistent SPD, physically guided optimization through the PGMF, RMS-based geometric refinement, and final validation. The implementation is carried out within \texttt{KrakenOS}, where optimization and evaluation of the design constraints are performed. A schematic representation of the computational workflow is shown in Fig.~\ref{fig:PhysandRMSFC}, while a detailed system layout is presented in a companion work on OPTICam-ARG\,\cite{najera2026opticaldesignopticamargthreechannel}. The implementation of the SPD construction and the two-stage optimization procedure is described in Sections~\ref{subSec:SPD}--\,\ref{subSec:StageII}.
        
    \subsection{Starting Point Design}\label{subSec:SPD}
    
        Within the multi-channel configuration, the SPD is constructed independently for each optical channel, allowing each focal reduction system to be optimized under equivalent physical constraints. Unlike conventional approaches based on heuristic initial configurations, which can limit convergence reliability in non-convex design spaces, the methodology defines the SPD within a constrained region of the parameter space. The initial configuration is guided by classical lens forms, such as the Cooke triplet, characterized by positive optical power in the outer elements and negative power in the central element, providing a balanced starting configuration. The SPD is first established through a paraxial description based on the ABCD matrix formalism \cite{10.1119/1.1970159}, which provides a global representation of the optical behavior. This formulation incorporates system-level constraints, including the target effective focal length and the paraxial focusing condition, together with a regularization term that penalizes extreme optical power values to promote realistic solutions. The resulting system is solved through numerical minimization of the residual formulation:
        \begin{equation}
        \begin{split}
        d\phi\,&=\,\sum_{i=1}^{3}\frac{\phi_{\mathrm{L}i}}{n_{\mathrm{L}i}}, \\
        d\mathrm{EFFL}\,&=\,\left| \mathrm{EFFL}{\mathrm{t}} - \mathrm{EFFL}{\mathrm{T}} \right|, \\
        d\,&=\,\mathrm{MS}[1,1],
        \end{split}
        \end{equation}
        where $\mathrm{MS}$ denotes the paraxial transfer matrix of the complete system.
        
        The paraxial solution is then extended to a third-order description, where optical powers are mapped to physically realizable geometric parameters of thick lens elements. The design variables are defined as the radii of curvature and a compensation distance enforcing the target effective focal length. The process is initialized from reference glasses with demonstrated performance \cite{castro+19}, such as S-FPL51 and F2HT, with the first and third lenses sharing the same material to reduce the dimensionality of the discrete search space while preserving the behavior of classical triplet configurations. The transition to third-order modeling is performed under the constraint $\mathrm{R}_{2i} = \beta\,\mathrm{R}_{1i}$, ensuring consistency with the paraxial optical power. This parametrization reduces the dimensionality of the design space while preserving the paraxial optical structure. The system is refined by minimizing residuals derived from third-order aberration theory,
        \begin{equation}
        \mathbf{f}(\mathbf{x})^{\mathsf{T}}\,=\,\bigl(
        S_{I},\,
        S_{II},\,
        S_{III},\,
        S_{IV},\,
        S_{V},\,
        C_{L},\,
        \Delta \mathrm{EFFL}
        \bigr),
        \end{equation}
        where $S_{I}$ through $S_{V}$ denote the Seidel aberration coefficients (spherical, coma, astigmatism, field curvature, and distortion), $C_{L}$ represents the longitudinal chromatic aberration, and $\Delta \mathrm{EFFL}$ represents the effective focal length constraint. Off-axis aberrations are evaluated at representative field positions, with $(0.00, 0.07)$ deg used to characterize coma and astigmatism. Chromatic aberration is evaluated at discrete wavelengths for each spectral channel: $(0.35, 0.43, 0.55)\,\mu$m (blue), $(0.55, 0.61, 0.67)\,\mu$m (green), and $(0.67, 0.81, 1.00)\,\mu$m (red). Convergence of the third-order residual formulation does not guarantee acceptance of the solution. Instead, the configuration is accepted only when both aberration balance and effective focal length constraints are satisfied; otherwise, the process returns to the third-order stage, reconstructing a new configuration under the imposed $\beta$ constraint while preserving the paraxial power distribution.
                
        Once a consistent configuration is obtained, the methodology enables structured exploration of the parameter space. Candidate materials are selected from the discrete $(n, V_d)$ space, restricting the search to optically similar materials within $\Delta n = 0.025$ and $\Delta V_d = 5.0$, and enforcing a minimum transmission threshold of $PT_{\mathrm{threshold}} = 0.8$ over $0.3$--$1.0\,\mu$m. To organize this exploration, candidate materials are ordered according to their proximity to the reference glass in the $(n,V_d)$ plane using a Euclidean distance metric. For each candidate, the optical parameters are optimized using a nonlinear LS solver implemented in \texttt{SciPy} (\texttt{least\_squares})\,\cite{2020SciPy-NMeth}, and the solutions are ranked according to the $EE_{50}$ radius, defined as the radius enclosing $50\%$ of the encircled energy, thereby prioritizing configurations with smaller polychromatic  $EE_{50}$ radii.
            
    \subsection{Stage I: Physically Guided Optimization}\label{subSec:StageI}
    
        The SPD obtained from the paraxial, third-order, and material exploration stages serves as the initial configuration for the physically guided optimization stage. Starting from this optically consistent configuration, the system is refined through minimization of the PGMF residuals (Eq.~\ref{eq:PGMF}), which constrain the accessible parameter space through aberration conditions directly linked to the optical behavior of the system prior to image-based refinement. The PGMF is evaluated through exact ray tracing within \texttt{KrakenOS} using a reduced set of representative rays. To extend the formulation to multiple field positions, the residual vector is expanded as
        \begin{equation} 
        \mathbf{f}(\mathbf{x})^{\mathsf{T}} = \bigl( \delta l_{F_C},\, w_r\,\delta r_{F_C},\, \delta C_{T,F_1},\, w_f\,\delta f_{F_1},\, \delta C_{T,F_2},\, w_f\,\delta f_{F_2},\, \Delta \mathrm{EFFL} \bigr), 
        \end{equation}
        where $F_C$ corresponds to the on-axis field, $F_1$ to the primary off-axis field defined in the SPD, and $F_2$ to an additional off-axis field at $(0.07, 0.07)$ deg. This formulation extends the PGMF to multiple field positions, ensuring consistency between constraints and the dimensionality of the design space. The residuals are evaluated across the spectral channels and representative wavelengths defined in the SPD. The system focus is defined at the $F_2$ position, corresponding to the edge-of-field condition, ensuring that the optimization prioritizes image quality at the field boundary.
                
        The system is refined by minimizing the objective function defined in Eq.~\ref{eq:Equation2OPT}. The solution is obtained iteratively by updating the system configuration based on the PGMF, with ray tracing and residual evaluation performed at each step. Convergence is achieved when the objective function stabilizes or a predefined performance criterion is met. Nonlinear LS optimization is used for local refinement, consistent with the SPD stage. In parallel, stochastic approaches such as Gaussian Monte Carlo sampling (MC), Genetic Algorithms (GA) implemented with \texttt{PyGAD}\,\cite{gad2023pygad}, and Artificial Bee Colony (ABC) using the \texttt{Bee Algorithm 1.0.2} library\,\cite{baronti2020analysis} enable exploration of neighboring regions of the design space around the local solution. The search region is restricted to perturbations within $\pm10\%$ of the optical parameters obtained from the LS solution. For the stochastic approaches, both GA and ABC are executed for 200 iterations. The GA implementation employs random parent selection, uniform crossover, and random mutation, while MC performs two iterations using Gaussian random realizations with 5000 samples each. This combination enables assessment of solution consistency under perturbations of the design variables within the neighborhood of the locally optimized configuration. The resulting configuration serves as the input for the subsequent refinement stage.
    
    \subsection{Stage II: Geometric Refinement}\label{subSec:StageII}

        Following the physically guided optimization, a second refinement phase is introduced to directly improve the geometric image quality at the detector plane. The RMS radius of the spot diagram is used as the objective function, evaluated across multiple field positions under the same spectral sampling defined previously. The field positions are selected as $(0.00,0.00)$, $(0.00,0.07)$, $(0.07,0.07)$, $(0.07,0.00)$, $(0.07,-0.07)$, and $(0.00,-0.07)$ deg, chosen to exploit rotational symmetry and provide representative coverage of the field of view for each channel. The pupil is sampled using a Gaussian quadrature distribution, providing an efficient representation of the system behavior with a reduced number of rays. The residual vector is defined as
        \begin{equation}
        \mathbf{f}_{\mathrm{RMS}}(\mathbf{x})^{\mathsf{T}} =
        \bigl(
        RMS_1,\,
        RMS_2,\,
        \ldots,\,
        RMS_n,\,
        \Delta \mathrm{EFFL}
        \bigr),
        \end{equation}
        where each term corresponds to the RMS value at a given field position. During this stage, the system focus is redefined at the $F_2$ position, preserving the edge-of-field criterion established in Stage~I and prioritizing image quality at the field boundary. The refinement is performed through nonlinear LS minimization, yielding the final \textit{Kraken solutions}, which are subsequently validated and fine-tuned within a commercial optical design environment.
    
    \subsection{Open and Reproducible Implementation}\label{subSec:Implementation}

\begin{figure}[ht]
            \centering
            \includegraphics[width=50mm]{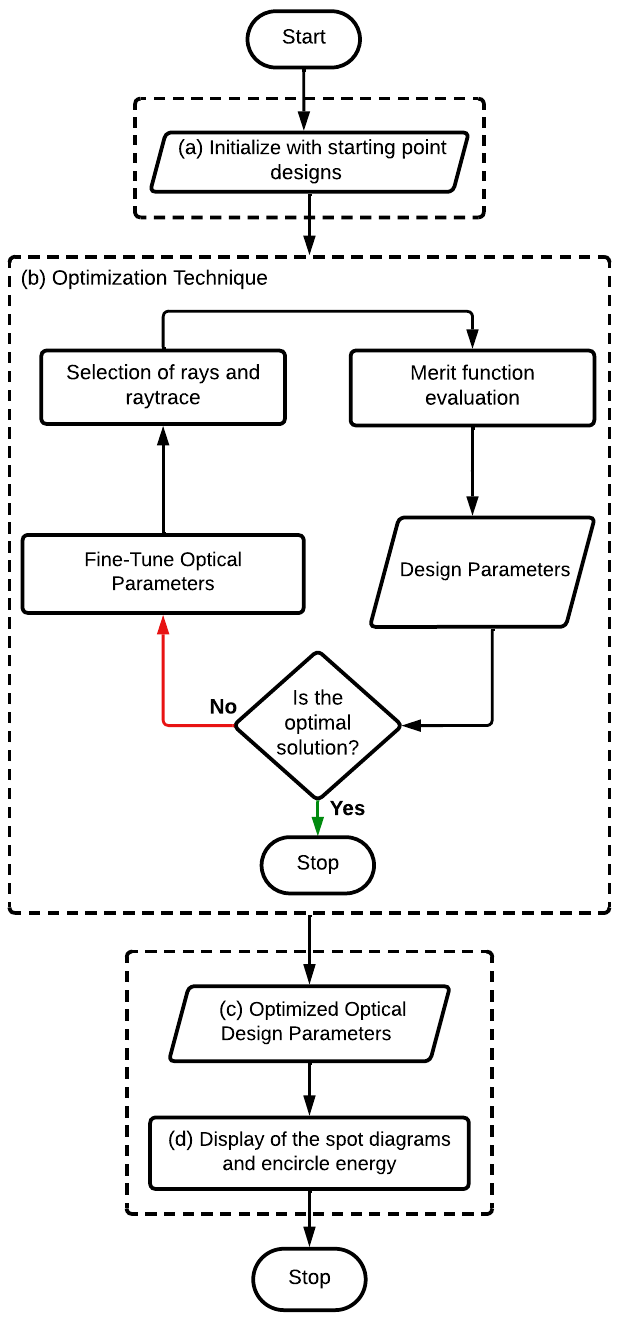}
            \caption{Workflow of the two-stage optimization procedure implemented within the \texttt{KrakenOS} framework.}
            \label{fig:PhysandRMSFC}
        \end{figure}

        The computational implementation directly follows the sequence defined by the SPD construction and the two-stage optimization strategy. A schematic overview of the methodology is shown in Fig.~\ref{fig:PhysandRMSFC}. The procedure is organized into four functional blocks: (i) system initialization from the SPD or a previous-stage solution; (ii) merit function evaluation through third-order models or exact ray tracing, depending on the optimization stage; (iii) iterative update of the design variables; and (iv) performance validation based on image quality metrics. These components form a unified computational pipeline in which the SPD provides the initial configuration for the subsequent optimization and refinement stages.
                
        The optical system is defined using an object-oriented structure in which each surface is parameterized by its geometry and material properties, and the complete system is assembled for sequential ray tracing. This structure enables direct access to each optical element, allowing explicit updates of design variables at every iteration. Parameters such as radii of curvature and distances are modified at the surface level through direct assignment, with updates first registered within the system data structure and subsequently applied to the optical model through a consolidation step. This separation ensures consistent propagation of changes throughout the optical model while enabling non-destructive evaluation of candidate solutions. Once a satisfactory solution is identified, the corresponding updates can be permanently applied. This modular structure supports a consistent transition from paraxial models to fully optimized configurations. In particular, \texttt{KrakenOS} provides tools for paraxial ray tracing and third-order aberration analysis through \texttt{Paraxial\_Cal} and \texttt{SeidelTool}, enabling the computation of the ABCD matrix and both total and individual Seidel aberrations.
                
        \texttt{KrakenOS} also incorporates optical glass catalogs from multiple providers, enabling direct access to refractive index and Abbe number. A dedicated tool (\texttt{GlassXtractor}) supports the structured material exploration described in Section~\ref{subSec:SPD}, integrating the discrete selection of optical materials into the computational workflow. For ray sampling, a pupil generation tool (\texttt{Pupil\_Cal}) defines marginal rays, chief rays, and structured distributions, including Gaussian quadrature sampling for RMS evaluation. For each field position, rays are generated from the entrance pupil, ensuring proper representation of on-axis and off-axis configurations. Ray intersections at the image plane are used to compute the residuals associated with both the PGMF and the RMS-based merit function. The resulting configurations are validated using spot diagrams and the $EE_{50}$ criterion. The implementation is fully reproducible, and the codes are publicly available in a dedicated GitHub repository(Code 1)\cite{Najera2026Kraken2Stage}, enabling independent verification and reuse of the methodology.

\section{Results}\label{Sec:Results}

    This section presents the optimized system and its corresponding optical performance. The material optimization process leads to a solution based on $\mathrm{K\text{-}PFK85}$–$\mathrm{ADF355}$–$\mathrm{K\text{-}PFK85}$, hereafter referred to as the KPFK-based system. The results are organized into two components. Section~\ref{subSec:KrakenSolutions} presents the resulting optical layout together with its image quality, while Section~\ref{subSec:PerformanceComparison} compares the optimization strategies used for local refinement and neighborhood exploration during Stage~I.

    \subsection{Kraken Solutions}\label{subSec:KrakenSolutions}

\begin{figure}[htbp]
            \centering
            \includegraphics[width=0.45\textwidth]{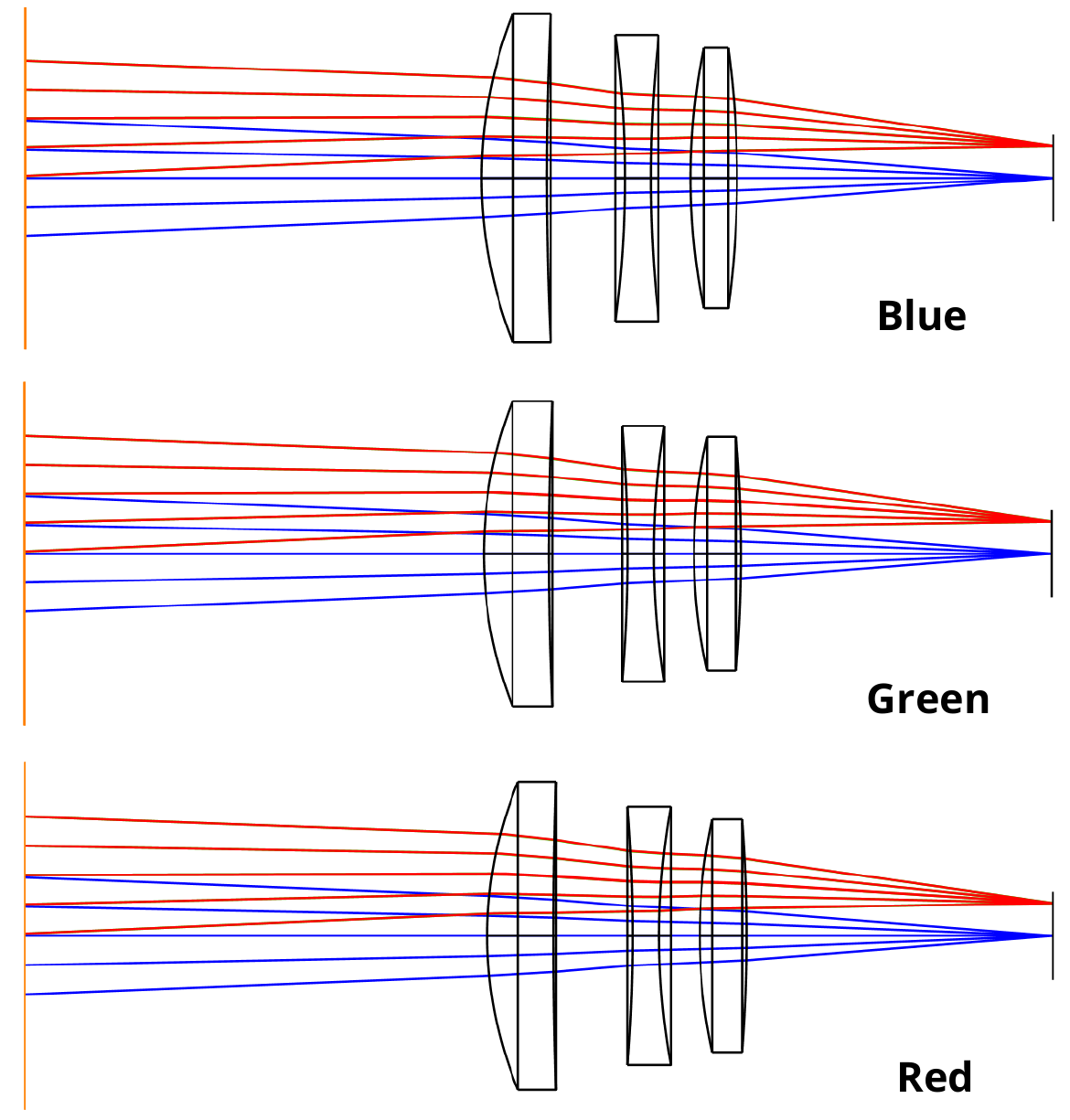}
            \caption{Optical layouts of the three focal reducer systems corresponding to the KPFK-based system for the blue, green, and red spectral channels. The layouts illustrate the final geometrical arrangement of the three-lens systems obtained from the proposed methodology prior to the inclusion of detector-side elements. The incident rays are defined from an orange reference plane introduced for visualization purposes, located $447.429$ mm from the primary mirror vertex and $157.887$ mm from the first surface of the first lens.}
            \label{fig:AllSystem}
        \end{figure}

        The resulting optical parameters for the KPFK-based system are summarized in Table~\ref{tab:Initial3LensConfigs}, which also includes the corresponding values for the reference material combination for completeness. The table lists the radii of curvature of the six optical surfaces and the corresponding compensation distance $d_{\mathrm{comp}}$ for each spectral channel. These values are defined prior to the inclusion of detector-side elements, such as filters and camera windows. For the KPFK-based system, the estimated distances between the last optical surface and the filter plane are $83.50$ mm, $80.96$ mm, and $78.64$ mm for the blue, green, and red channels, respectively. These values are in close agreement with the final optimized distances reported in the companion work ($83.29$ mm, $80.96$ mm, and $79.23$ mm), indicating that the global geometry of the optical system is already well established. The corresponding bending parameters used during the third-order construction are $\beta_1 = 6.0$, $8.5$, and $11.75$ for the blue, green, and red channels, respectively, while $\beta_2=-0.5$ and $\beta_3=-3$ are kept fixed across all channels. Figure~\ref{fig:AllSystem} illustrates the corresponding optical layouts for the three spectral channels obtained from the proposed framework.
        
        \begin{table}[htbp]
            \caption{Kraken solutions for the three-lens focal reducer system. Radii of curvature ($R_1$–$R_6$) and compensation distance $d_{\mathrm{comp}}$ for each spectral channel.}
            \label{tab:Initial3LensConfigs}
            \centering
            \small
            \begin{tabular}{cccccccc}
            \hline
            Channel & $R_1$ & $R_2$ & $R_3$ & $R_4$ & $R_5$ & $R_6$ & $d_{\mathrm{comp}}$ \\
             & (mm) & (mm) & (mm) & (mm) & (mm) & (mm) & (mm) \\
            \hline
            
            \multicolumn{8}{c}{$\mathrm{K\text{-}PFK85}$, $\mathrm{ADF355}$, $\mathrm{K\text{-}PFK85}$} \\
            \hline
            Blue  & 153.58 & 1221.04 & -382.57 & 447.27 & 227.82 & -323.35 & 109.50 \\
            Green & 143.68 & 1173.81 & -533.91 & 279.66 & 182.45 & -473.48 & 106.96 \\
            Red   & 137.18 & 1609.07 & -559.52 & 245.66 & 183.80 & -516.67 & 104.64 \\
            
            \hline
            
            \multicolumn{8}{c}{$\mathrm{S\text{-}FPL51}$, $\mathrm{F2HT}$, $\mathrm{S\text{-}FPL51}$} \\
            \hline
            Blue  & 152.11 & 499.69  & -624.78 & 431.95 & 194.27 & -577.35 & 110.51 \\
            Green & 145.58 & 776.54  & -636.65 & 303.69 & 198.10 & -556.12 & 107.22 \\
            Red   & 145.86 & 1791.37 & -360.71 & 335.26 & 248.57 & -312.55 & 106.93 \\
            
            \hline
            \end{tabular}
        \end{table}
        
\begin{figure}[ht]
        \centering
        \includegraphics[width=0.57\textwidth]{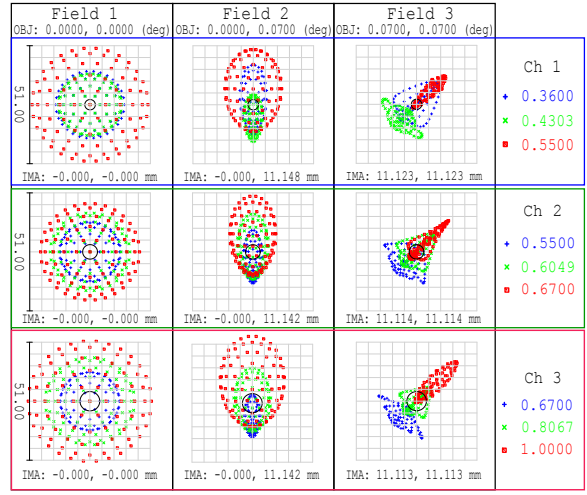}
        \caption{Spot diagrams for the final three-lens system using $\mathrm{K\text{-}PFK85}$, $\mathrm{ADF355}$, and $\mathrm{K\text{-}PFK85}$. For each spectral channel (rows), three field positions are shown in the image plane: the on-axis field $(0.00,\,0.00)$ and the off-axis fields $(0.00,\,0.07)$ and $(0.07,\,0.07)$. Colors indicate the wavelengths considered in each channel. The scale bar corresponds to $51\,\mu$m and is included for visual reference.}
        \label{fig:3lens_spot}
        \end{figure}

        The image quality of the Kraken solutions is characterized through the analysis of spot diagrams. Figure~\ref{fig:3lens_spot} shows the spot diagrams for the three spectral channels of the instrument, arranged by rows: channel\,1 (top), channel\,2 (middle), and channel\,3 (bottom). For each channel, one on-axis field and two off-axis positions, corresponding to the fields $F_1$ and $F_2$ defined in Section~\ref{subSec:StageI}, are included. All spot diagrams are displayed using a common scale bar of $51\,\mu$m, corresponding to a seeing of $\sim1.15''$ for the plate scale of $22.6\,''\,\mathrm{mm}^{-1}$, and are shown for visual comparison. The use of color allows identification of the different wavelengths considered in each channel, while the corresponding positions in the object and image planes are indicated for each diagram. To quantify the spatial extent of the spot distributions, the geometric spot radius (GEO) is defined as the maximum radial distance of the traced rays from the centroid of the spot diagram, representing the outer extent of the ray distribution in the image plane. For the KPFK-based system, the GEO radii lie in the range of $17.17\,\mu$m to $28.90\,\mu$m, while the RMS radii remain below $17\,\mu$m across all evaluated fields. The corresponding RMS values are $15.34\,\mu$m, $11.56\,\mu$m, and $9.77\,\mu$m for the blue channel; $14.57\,\mu$m, $10.31\,\mu$m, and $8.41\,\mu$m for the green channel; and $16.78\,\mu$m, $12.00\,\mu$m, and $10.30\,\mu$m for the red channel, evaluated at $(0.00,0.00)$, $(0.00,0.07)$, and $(0.07,0.07)$, respectively.
        
\begin{figure}[ht!]
        \centering
        \includegraphics[width=0.68\textwidth]{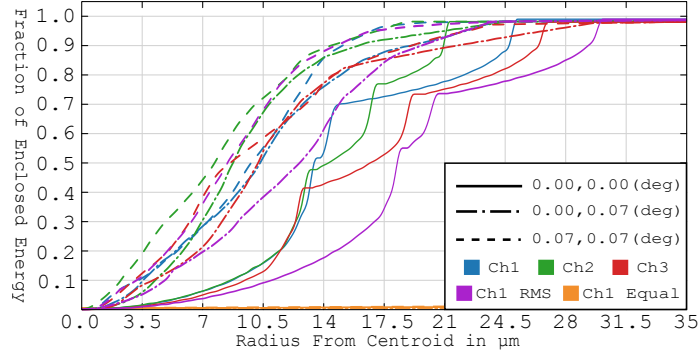}
        \caption{Encircled geometric energy curves for the three spectral channels: blue (channel\,1), green (channel\,2), and red (channel\,3). For channel\,1, the purple curves (Ch1 RMS) show direct RMS refinement from the SPD, bypassing the intermediate PGMF stage, whereas the orange curves (Ch1 Equal) show the corresponding workflow initialized from an equal-pair configuration. The curves are normalized to the diffraction limit. Line styles indicate field positions: solid (on-axis), dash-dotted (off-axis), and dashed (diagonal).}
        \label{fig:3lens_ee}
        \end{figure}

        The energy concentration is characterized through the $EE_{50}$ metric, computed polychromatically. Figure~\ref{fig:3lens_ee} shows the encircled geometric energy curves for the three spectral channels, distinguished by representative colors, evaluated at the same field positions and normalized to the diffraction limit. The horizontal axis represents the EE fraction, while the vertical axis indicates the radius from the spot-diagram centroid in $\mu$m. For the KPFK-based system, the on-axis $EE_{50}$ radii are $13.58\,\mu$m, $14.51\,\mu$m, and $16.7\,\mu$m for the blue, green, and red channels, respectively. These on-axis values provide a conservative estimate of the image quality across the field, as they correspond to the largest $EE_{50}$ radii among the evaluated field positions. Using the plate scale of $22.6\,''\,\mathrm{mm}^{-1}$, the corresponding angular radii are approximately $0.31''$, $0.33''$, and $0.38''$ for the blue, green, and red channels, respectively, remaining smaller than the typical site seeing.
        
        To illustrate the contribution of the intermediate optimization stage, Fig.~\ref{fig:3lens_ee} also includes the blue-channel solution obtained by applying RMS refinement directly to the third-order starting point, without Stage~I. The complete workflow reduces the on-axis $EE_{50}$ radius from $18.21\,\mu$m to $13.58\,\mu$m and the intermediate off-axis value from $12.54\,\mu$m to $11.56\,\mu$m, while both approaches produce comparable energy concentration for the diagonal field. Additionally, the complete workflow requires $40.71$\,s, compared with $64.80$\,s for direct RMS refinement, corresponding to an overall computational time reduction of approximately $37\%$. To examine the role the physically grounded SPD, Fig.~\ref{fig:3lens_ee} also includes the workflow initialized from an equal-pair configuration ($R_1=R_2$, $R_3=R_4$, and $R_5=R_6$). Unlike the SPD initialization used in the proposed workflow, this equal-pair initialization converged to poorer solutions, with on-axis $EE_{50}$ radii exceeding $260\,\mu$m and best-case values above $170\,\mu$m. Furthermore, the resulting designs did not preserve the target focal reduction.
        
         A comparison with the final optimized design, shown in Table~\ref{tab:radius_deviation}, reveals that the largest variations occur in the internal radii of curvature, particularly in $R_2$, while the outer surfaces remain comparatively stable. This behavior reflects the role of the internal surfaces in the final balancing of aberrations during the image-quality refinement stage. To characterize the relationship between the \texttt{KrakenOS} solution and the final \textit{ZEMAX}\textsuperscript{\textregistered} design, a first-order optical analysis was performed. The equivalent optical power of the complete three-lens group differs by less than $3\%$ for all three spectral channels, while the maximum variation in the optical power of an individual lens is approximately $25\%$. Consequently, the differences in the effective paraxial image-plane position remain below $1$ mm for all channels.
        
        \begin{table}[htbp]
        \caption{Differences between the initial radii obtained with \texttt{KrakenOS} and the final optimized values in \textit{ZEMAX}\textsuperscript{\textregistered} OpticStudio for the $\mathrm{K\text{-}PFK85}$–$\mathrm{ADF355}$–$\mathrm{K\text{-}PFK85}$ configuration. Differences are defined as $\Delta R_i = R_i^{\mathrm{final}} - R_i^{\mathrm{init}}$. The difference in the effective distance to the filter plane is also included.}
        \label{tab:radius_deviation}
        \centering
        \small
        \begin{tabular}{cccccccc}
        \hline
        Channel & $\Delta R_1$ & $\Delta R_2$ & $\Delta R_3$ & $\Delta R_4$ & $\Delta R_5$ & $\Delta R_6$ & $\Delta d$ \\
         & (mm) & (mm) & (mm) & (mm) & (mm) & (mm) & (mm) \\
        \hline
        
        Blue  & -6.90  & -247.97   & -347.71 & -174.22 & -53.73 & -288.07 & +0.21 \\
        Green & -6.03  & 15626.19  & 119.52  & -40.50  & 12.55  & 91.64   & 0.00 \\
        Red   & -2.32  & -12109.07 & 154.39  & -38.74  & -9.38  & 121.54  & -0.59 \\
        
        \hline
        \end{tabular}
        \end{table}

    \subsection{Performance Comparison}\label{subSec:PerformanceComparison}
    
\begin{figure}[ht!]
        \centering
        \includegraphics[width=0.7\textwidth]{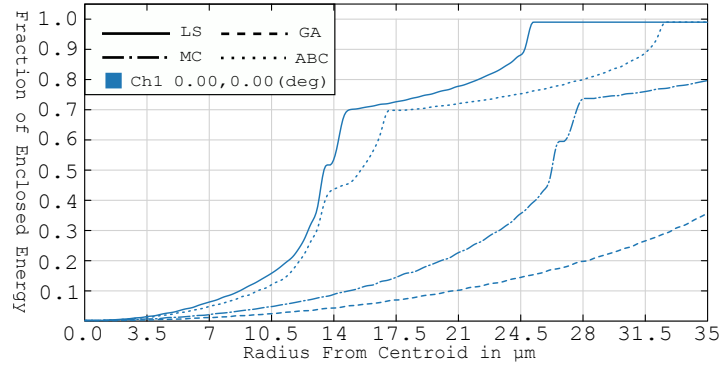}
        \caption{Encircled geometric energy curves for the blue channel at the on-axis field obtained with different optimization strategies. The LS solution is shown with a solid line, MC with a dash-dotted line, GA with a dashed line, and ABC with a dotted line.}
        \label{fig:EE_Methods}
        \end{figure}

        The performance of different optimization strategies within the proposed framework is evaluated using consistent performance metrics. The comparison includes nonlinear LS optimization together with stochastic approaches such as MC, GA, and ABC, which explore neighboring regions of the design space around the local solution obtained during Stage~I. The blue channel is used as a representative case due to its stronger sensitivity to chromatic and off-axis aberrations. Table~\ref{tab:performance_comparison} summarizes the resulting RMS and $EE_{50}$ values for the on-axis field, which provides a representative and conservative estimate of the system performance, as discussed in the previous section. Among the stochastic strategies explored, ABC produces the closest solution to the LS result, resulting in RMS and $EE_{50}$ values of $18.52\,\mu$m and $15.49\,\mu$m, respectively. MC achieves intermediate performance with values of $26.53\,\mu$m and $26.24\,\mu$m, while GA produces the largest deviations, reaching $38.88\,\mu$m and $38.01\,\mu$m. These results reflect the sensitivity of stochastic exploration strategies to the sampling conditions and exploration parameters within the non-convex design space. Nevertheless, all strategies remain constrained around the optically guided local solution obtained during Stage~I. Figure~\ref{fig:EE_Methods} shows the corresponding encircled energy curves for the blue channel at the on-axis field, where the LS solution is represented by a solid line, MC by a dash-dotted line, GA by a dashed line, and ABC by a dotted line. The figure illustrates the closer agreement between the LS and ABC solutions, particularly in the low-radius region associated with higher energy concentration near the centroid.

        \begin{table}[htbp]
        \caption{Performance comparison of optimization strategies for the blue channel. Reported values correspond to the on-axis field.}
        \label{tab:performance_comparison}
        \centering
        \small
        \begin{tabular}{cccc}
        \hline
        Method & RMS ($\mu$m) & EE$_{50}$ ($\mu$m) & EE$_{50}$ (arcsec) \\
        \hline
        LS  & 15.34 & 13.58 & 0.31 \\
        MC  & 26.53 & 26.24 & 0.59 \\
        GA  & 38.88 & 38.01 & 0.86 \\
        ABC & 18.52 & 15.49 & 0.35 \\
        \hline
        \end{tabular}
        \end{table}
        
\section{Discussion}\label{Sec:Discussion}

    The results demonstrate that the proposed two-stage framework provides a consistent and physically interpretable pathway toward high-quality optical solutions. By combining a physically guided merit function in Stage~I with geometric refinement in Stage~II, the methodology balances aberration control and energy concentration throughout the optimization process. A direct comparison with RMS refinement applied to the same third-order starting configuration shows that the intermediate PGMF improves energy concentration before the final image-quality optimization, particularly for the most demanding on-axis field. Likewise, the equal-pair initialization experiment demonstrates the importance of the SPD, since replacing it with an equal-pair starting configuration leads to substantially poorer image quality and fails to preserve the prescribed focal reduction. Together, these experiments show that the SPD and the intermediate PGMF stage play complementary roles in the proposed workflow. The SPD provides a starting configuration that preserves the prescribed first-order behavior, whereas the PGMF stage provides an intermediate solution that leads to improved subsequent RMS refinement. In addition, the final optimized systems operate in a regime where the residual optical aberrations remain below the image degradation associated with the representative atmospheric seeing conditions considered in the design. A first-order optical analysis was performed to characterize the relationship between the \texttt{KrakenOS} solution and the final \textit{ZEMAX}\textsuperscript{\textregistered} design. The equivalent optical power of the complete three-lens group and the paraxial image-plane position remain nearly unchanged, while the individual lens powers exhibit a moderate variations. These results indicate that the global paraxial characteristics of the optical system are preserved during the final refinement.

    The observed differences between stochastic strategies are influenced by the exploration parameters and computational budget. In particular, the GA implementation employs random parent selection together with uniform crossover and random mutation, increasing the stochastic dispersion of the explored solutions. Similarly, the MC approach relies on a limited number of Gaussian random realizations, restricting the sampling density within the local neighborhood of the LS solution. Although increasing the number of realizations or iterations could potentially yield solutions with lower RMS and $EE_{50}$ values, such modifications would also increase the computational cost associated with repeated ray-tracing evaluations. These observations reflect the different ways in which each stochastic strategy explores the local neighborhood defined by the LS solution obtained in Stage~I. In this sense, the methodology can be interpreted as a structured reduction of the effective dimensionality of the problem, where the SPD defines an optically meaningful starting region, Stage~I identifies a physically consistent solution within that region, and the subsequent stochastic exploration is focused on its local refinement rather than on unconstrained global exploration.
    
    The present work demonstrates the methodology using a three-element focal reducer with a Cooke-triplet-like power distribution; however, the proposed framework is not intrinsically restricted to this particular architecture. Its underlying philosophy consists of constructing physically meaningful residuals from paraxial and third-order optical models and incorporating them into a structured optimization workflow. Consequently, the same framework can, in principle, be extended to other refractive systems with a larger number of optical elements as well as to catadioptric configurations. Despite these advantages, the current implementation presents limitations. In particular, the construction of the SPD and the associated merit functions depends on a predefined system architecture. Extending the methodology to a different optical configuration requires identifying an appropriate optical architecture for the target application, selecting optical materials, and redefining the corresponding paraxial power distribution that defines the initial design region. The associated first-order and third-order models must then be reformulated so that the number of unknowns and physically meaningful constraints remain consistent, followed by the construction of new physically grounded residual merit functions together with the corresponding RMS-based refinement strategy. Consequently, the methodology is not yet fully automated in terms of generating SPDs or constructing merit functions for arbitrary optical configurations. Nevertheless, this limitation does not affect the validity of the framework within the design regime explored. Instead, it identifies a natural direction for future research, namely the development of automated strategies capable of generating physically grounded SPDs and associated merit functions for arbitrary optical architectures.

\section{Conclusions} \label{Sec:Conclusions}

    This work does not seek to replace modern optical design principles, but rather to formalize classical design strategies within a reproducible numerical environment. The proposed two-stage approach integrates paraxial modeling, third-order aberration balancing, exact ray tracing, and RMS-based image refinement into a unified computational pipeline. Starting from optically consistent SPD, the methodology combines a physically grounded merit function in Stage~I with RMS-based refinement in Stage~II, enabling structured exploration of the optical design space.
    
    The results show that the method produces optically consistent solutions that closely approximate the final optimized system while reducing reliance on heuristic initial conditions. By embedding aberration-based constraints directly into the refinement procedure, the methodology guides the exploration toward viable regions of the parameter space prior to image-based optimization. Additional validation experiments support the proposed framework by demonstrating the complementary contributions of the PGMF and the SPD to image quality, preservation of the prescribed first-order optical characteristics, and the physical interpretability of the optimization process. The comparison between local and stochastic strategies demonstrates that the proposed approach provides consistent starting points for neighborhood exploration, while illustrating how different exploration mechanisms influence energy concentration and solution variability across the evaluated configurations. The results also highlight the influence of exploration parameters and computational budget on the behavior of stochastic approaches, emphasizing the tradeoff between sampling density and computational cost in ray-tracing-based optimization. Overall, the proposed framework provides a reproducible and physically interpretable pathway from classical optical modeling to numerical refinement. While the current implementation requires predefined system architectures, future developments aimed at automating the generation of SPDs and merit functions are expected to extend the applicability of the methodology to more general optical configurations.

\begin{backmatter}

    \bmsection{Funding}

        This work was supported by DGAPA--PAPIIT (IT102125). M.R.N. also acknowledges the Secretaría de Ciencia, Humanidades, Tecnología e Innovación (SECIHTI) for financial support through a doctoral scholarship (CVU 1044690).
    
    \bmsection{Acknowledgment}
    
        The authors acknowledge the use of the following software packages: \texttt{KrakenOS}~\cite{herrera2022krakenos}, \texttt{NumPy}~\cite{harris2020array}, \texttt{PyGAD}~\cite{gad2023pygad}, \texttt{Bee-Algorithm}~\cite{baronti2020analysis}, \texttt{Matplotlib}~\cite{hunter2007matplotlib}, \texttt{SciPy}~\cite{2020SciPy-NMeth}, \texttt{PyVista}~\cite{sullivan2019pyvista}, and \texttt{VTK}~\cite{vtkBook}. The authors also acknowledge the use of ChatGPT~\cite{OpenAI_ChatGPT} for assistance with language editing and grammar revision
    
    \bmsection{Disclosures}
    
        The authors declare no conflicts of interest.
    
    \bmsection{Data availability}
    
        The source code used to generate the results presented in this paper is publicly available in Ref.~\cite{Najera2026Kraken2Stage} (Code~1). The numerical data underlying the results presented in this paper are available from the corresponding author upon reasonable request.

\end{backmatter}

\bibliography{RFRs}
\end{document}